# Symbolic regression based genetic approximations of the Colebrook equation for flow friction

Pavel Praks [1,2,*] and Dejan Brkić [1,*]

[1] European Commission, Joint Research Centre (JRC), Directorate C: Energy, Transport and Climate, Unit C3: Energy Security, Distribution and Markets, Via Enrico Fermi 2749, 21027 Ispra (VA), Italy

[2] IT4Innovations National Supercomputing Center, VŠB - Technical University of Ostrava, 17. listopadu 2172/15, 708 00 Ostrava, Czech Republic

* Correspondence: Pavel.Praks@ec.europa.eu, Pavel.Praks@vsb.cz (P.P.); dejanbrkic0611@gmail.com (D.B.)

**Abstract:** Widely used in hydraulics, the Colebrook equation for flow friction relates implicitly to the input parameters; the Reynolds number, Re and the relative roughness of inner pipe surface, ε/D with the output unknown parameter; the flow friction factor, $\lambda$; $\lambda=f(\lambda, Re, \varepsilon/D)$. In this paper, a few explicit approximations to the Colebrook equation; $\lambda \approx f(Re, \varepsilon/D)$, are generated using the ability of artificial intelligence to make inner patterns to connect input and output parameters in explicit way not knowing their nature or the physical law that connects them, but only knowing raw numbers, {Re, ε/D}→{λ}. The fact that the used genetic programming tool does not know the structure of the Colebrook equation which is based on computationally expensive logarithmic law, is used to obtain better structure of the approximations which is less demanding for calculation but also enough accurate. All generated approximations are with low computational cost because they contain a limited number of logarithmic forms used although for normalization of input parameters or for acceleration, but they are also sufficiently accurate. The relative error regarding the friction factor $\lambda$, in best case is up to 0.13% with only two logarithmic forms used. As the second logarithm can be accurately approximated by the Padé approximation, practically the same error is obtained also using only one logarithm.

---

**1. Introduction**

Colebrook equation for flow friction is one of the most used formulas in hydraulics, which is a branch of civil engineering, that deals with the conveyance of liquids through pipes. It is also widely used in mechanical, petroleum and chemical engineering, etc., wherever flow through pipes occur. It is empirical relation developed by Colebrook [1] based on his experiment with White [2]. The experiment dealt with flow of air/liquid through artificially roughened pipes; Eq. (1):

$$\frac{1}{\sqrt{\lambda}} = -2 \cdot \log_{10}\left(\frac{2.51}{Re} \cdot \frac{1}{\sqrt{\lambda}} + \frac{\varepsilon}{3.71 \cdot D}\right) \qquad (1)$$

In Eq. (1), $\lambda$ is the Darcy flow friction factor, Re is the Reynolds number, and ε/D is the relative roughness of inner pipe surface (all three quantities are dimensionless).

In the Colebrook equation, the flow friction factor $\lambda$ is implicitly given, $\lambda=f(\lambda, Re, \varepsilon/D)$ where it can be expressed in an explicit way only approximately [3-8], $\lambda \approx f(Re, \varepsilon/D)$ or otherwise the original equation can be solved iteratively [9, 10]. Today, it is important not only to have accurate, but also computationally efficient approximations [11-13]. Here, we used the ability of artificial intelligence to connect input data; in our case the Reynolds number, Re and the relative roughness of inner pipe surface, ε/D with the output parameter; in our case the flow friction factor, $\lambda$ not knowing the structure of the Colebrook equation [14-17]. So, we used the ability of artificial intelligence to connect input with output data to form patterns not knowing the nature of the data or physical law that connects them (similar approach is valid also for other branch of hydraulics [18, 19]). In that way, we tried to avoid the computationally expensive logarithmic law on which the Colebrook equation is based. As a final product we developed few low-cost but very accurate explicit approximations to the Colebrook equation.

## 2. Methods used, preparation of data and software tool, results, structure of approximations, accuracy and comparative analysis

The main idea is to use the ability of artificial intelligence to connect input data sets; in our case the Reynolds number, Re and the relative roughness of inner pipe surface, ε/D with the output data set; in our case the flow friction factor, $\lambda$; {Re, ε/D}→{$\lambda$}, not knowing the physical law which connects input to output. Sign "→" practically represents the Colebrook equation; Eq. (1), but the genetic programming tool is not aware of that fact. To prepare data to feed the genetic programming tool, we covered the whole practical domain of applicability of the Colebrook equation; which is for the Reynolds number, Re between 4000 and $10^8$ (whole turbulent flow covered) and for the relative roughness of inner pipe surface, ε/D up to 0.05 (pipe covered from practically smooth to the very rough) [20] with a mesh which consists of 90 thousand intersection points {Re, ε/D} for which we calculated very accurately the flow friction factor, $\lambda$ using the Colebrook equation; Eq. (1); Figure 1:

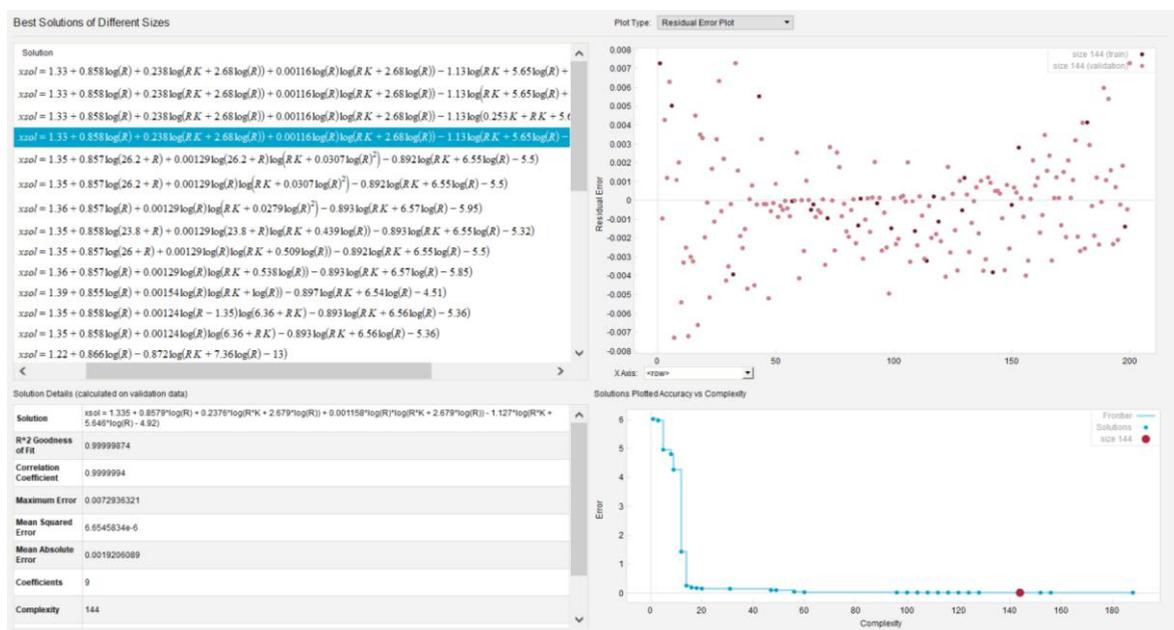

**Figure 1.** An example of Eureqa [computer software] interface: Best approximate solutions of the Colebrook equation with various complexity, which were automatically found by Eureqa, are on the left, whereas the residual error plot of a selected analytical model for 200 pairs together with an accuracy vs complexity plot of solutions is on the right

Having 90 thousand combinations; in order to test robustness of the symbolic regression algorithms we fed the genetic programming tool with 200 triplets {Re, ε/D}$_i$, {$\lambda$}$_i$, hoping that it will connect {Re, ε/D}$_i$→{$\lambda$}$_i$ accurately. The input sample was generated according to the uniform density function of each input variable. The low-discrepancy Sobol sequences were employed [21]. These so called quasirandom sequences have useful properties. In contrary to random numbers, quasirandom numbers covers the space more quickly and evenly. Thus, they leave very few holes. We used [computer software] Eureqa by Nutonian, Inc., Boston, MA, as a genetic programming tool [22, 23]. The symbolic regression approach adopted herein [24-29] is based upon genetic programming wherein a population of functions is allowed to breed and mutate with the genetic propagation into subsequent generations based upon a survival-of-the-fittest criteria [30]. The main goal of this study is to make accurate and computationally cheap explicit approximations of the Colebrook equation, where computationally cheap means to contain the least possible number of logarithmic functions and non-integer powers [31-36].

We can see that approximations found by Eureqa [computer software] have form $\{R, K\} \rightarrow xsol$, where the symbol $R$ denotes the Reynolds number, $K$ represents relative roughness and $xsol = \frac{1}{\sqrt{\lambda}}$. All accurate models are computationally expensive, as they contain many logarithmic terms with different arguments. Thus, we can see that Eureqa itself requires human knowledge, in order to obtain an accurate but still a computationally cheap approximation of the Colebrook equation. Consequently, we will combine several approaches in this paper: Eureqa [computer software] [22, 23], the fixed-point iteration [9] and Padé approximation [10]. According to our numerical experiments, Eureqa seems to be useful especially for finding a computationally cheap rational approximation of the Colebrook solution, which serves as a good starting point for the fixed-point iteration method (acceleration). Finally, Padé approximation is used as a cheap but very accurate approximation of the logarithm in the second and the successful iterations of the fixed-point method.

*2.1. Input parameters in their raw form*

Using the input parameters in their raw form $\{Re, \varepsilon/D\}_i \rightarrow \{\lambda\}_i$, Eureqa, the used genetic programming tool gives a set of approximations in polynomial forms [12]. Knowing that logarithmic expressions and non-integer powers are expensive for computation, we hoped that we have fully accomplished our task. Unfortunately, Eureqa gives a number of not very accurate solutions and here we show Eq. (2) with the relative error of $\lambda_0$ even up to 16.56% in respect to the accurate $\lambda$, where the relative error [5, 26] is defined as $(|\lambda_{accurate}-\lambda|/\lambda_{accurate}) \cdot 100\%$, where $\lambda_{accurate}$ is calculated in iterative procedure using the original implicitly given Colebrook equation [6, 9]; Eq. (1), while $\lambda$ is obtained through the presented approximations; Eqs. (2-6). In Eq. (2), "↔" means related but not sufficiently accurate:

$$\frac{1}{\sqrt{\lambda_0}} \leftrightarrow \frac{4.34 \cdot Re}{Re + 129000 \cdot Re \cdot \frac{\varepsilon}{D} + 7850000} + \frac{781 \cdot Re}{187 \cdot Re + 133000 \cdot Re \cdot \frac{\varepsilon}{D} + 8960000} - 20.5 \cdot \frac{\varepsilon}{D} + 4.85 \quad (2)$$

On the other hand, we found that the accuracy can increase significantly using one fixed-point iterative cycle of acceleration [9]; Eq. (2a), after which accuracy of $\lambda_1$ increases up to 0.98%.

$$\left. \begin{array}{c} \frac{1}{\sqrt{\lambda_1}} \approx -2 \cdot log_{10}(y_1) \\ \vdots \\ \frac{1}{\sqrt{\lambda_{i+1}}} \approx -2 \cdot log_{10}(y_{i+1}) \end{array} \right\} \quad (2a)$$

In Eq. (2a), "≈" means reasonably accurate enough and arguments of logarithms are defined by; Eq. (2b):

$$\left. \begin{array}{c} y_1 \approx \frac{2.51}{Re} \cdot \underbrace{\left( \begin{array}{c} \frac{4.34 \cdot Re}{Re + 129000 \cdot Re \cdot \frac{\varepsilon}{D} + 7850000} + \\ + \frac{781 \cdot Re}{187 \cdot Re + 133000 \cdot Re \cdot \frac{\varepsilon}{D} + 8960000} - 20.5 \cdot \frac{\varepsilon}{D} + 4.85 \end{array} \right)}_{\frac{1}{\sqrt{\lambda_0}}} + \frac{\varepsilon}{3.71 \cdot D} \\ \vdots \\ y_{i+1} \approx \left( \frac{2.51}{Re} \cdot \frac{1}{\sqrt{\lambda_i}} + \frac{\varepsilon}{3.71 \cdot D} \right) \end{array} \right\} \quad (2b)$$

The simple fixed-point iterative procedure [6, 9]; Eq. (2a) in case of the Colebrook equation is fast; $\lambda_0 \rightarrow 16.56\%$, $\lambda_1 \rightarrow 0.98\%$, $\lambda_2 \rightarrow 0.13\%$, etc. (Figure 2). Thus, using only two logarithmic forms high accuracy of $\lambda_2 \rightarrow 0.13\%$ is reached. Results are in the form $\{\lambda\}_0 \leftrightarrow \{Re, \varepsilon/D\}_0$, $\{\lambda\}_1 \approx \{log_{10}(\lambda_0)\}_1$, $\{\lambda\}_2 \approx \{log_{10}(log_{10}(\lambda_0))\}_2$, etc, where "↔" means related but not sufficiently accurate, while "≈" reasonably accurate enough. This approach with acceleration is widely used in development of approximations of the Colebrook equation [37-42]. The error can be further reduced by using one more accelerating step as shown, or using genetic algorithms [25, 29, 36], Excel fitting tool [27] or the Monte Carlo method [43, 44].

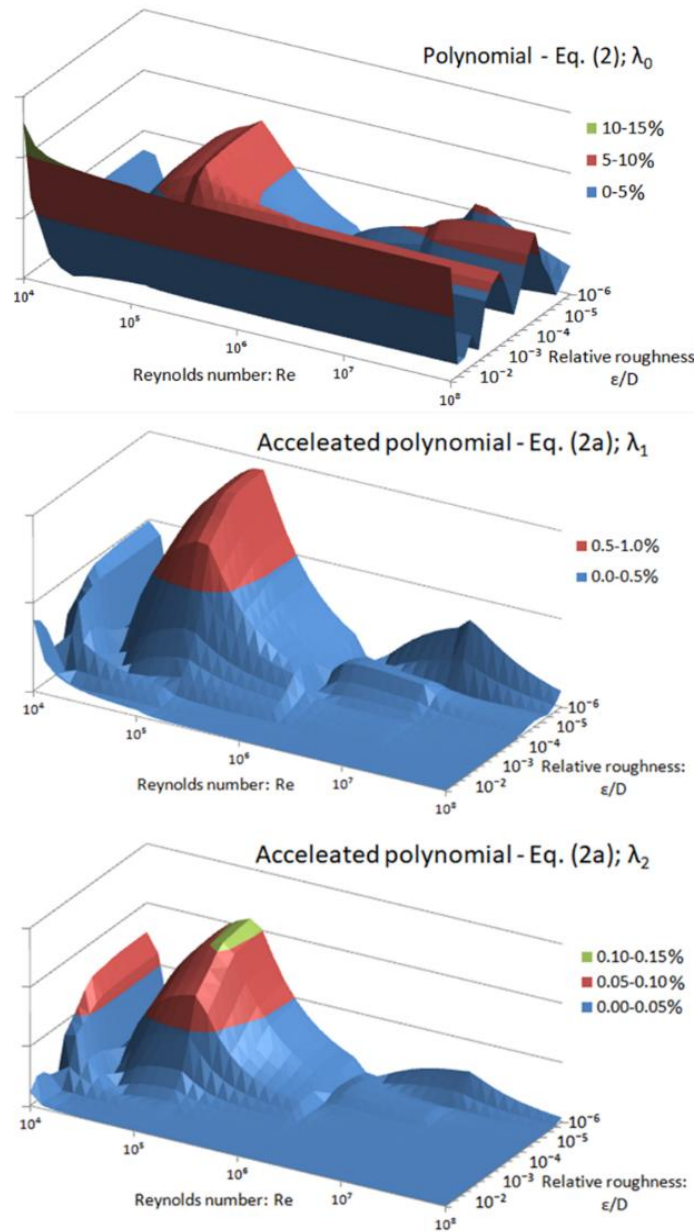

**Figure 2.** Distribution of the relative error of λ over the domain of applicability of the Colebrook equation introduced by polynomial Eq. (2) – up, Eq. (2a) after first step of fixed-point acceleration – middle, and the second step of acceleration - down; relative error up to 16.56%, up to 0.98% and up to 0.13% respectively

*2.2. Normalized input parameters*

Unfortunately, in our case using the input parameters in their raw form the accuracy was not at a high level without acceleration, so having previous experience with the same problem where we used Artificial Neural Network [15, 16] to simulate results, we normalized parameters $a=\log_{10}(Re)$, $b=-\log_{10}(\varepsilon/D)$, in order to avoid discrepancy in the scale which are in raw form $1000<Re<10^8$ and $\varepsilon/D \ll 1$ and after normalization $3.5<a<8$ and $1.3<b<6.5$ (Eureqa, software used a as genetic programming tool also suggested to us a data normalization process) [34-36]. The normalization gives relatively good results, and genetic programming tool generated more accurate results without knowing that the logarithmic form of the Colebrook equation was originally used but only knowing the predicted input and output datasets; Figure 3:

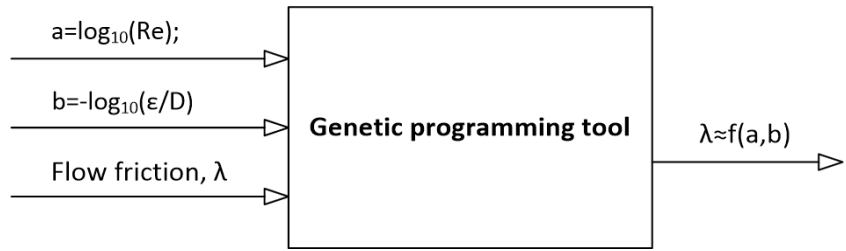

**Figure 3.** Genetic programming tool makes λ≈f(a,b) without knowing the physical law that connects a and b

Using our previous experience [15] with the training of the Artificial Neural Network where very good results were achieved through the normalization of parameters; a=log$_{10}$(Re), b=-log$_{10}$(ε/D), the genetic programming tool generated a dozen equations with different levels of accuracy and complexity, but fortunately none of them contain logarithms or non-integer power terms. Here we present the four most successful explicit approximations; Eqs. (3-6). Adding one additional logarithmic form for acceleration using one additional fixed-point iterative step [9, 45]; Eq. (2a), accuracy of the approximations increase significantly (about 10 times); Eqs. (3a-6a). Results are in the form {λ}$_0$↔{a=log$_{10}$(Re), b=-log$_{10}$(ε/D)}$_0$, {λ}$_1$≈{log$_{10}$(λ$_0$)}$_1$, {λ}$_2$≈{log$_{10}$(log$_{10}$(λ$_0$))}$_2$, etc, where "↔" means related but not sufficiently accurate, while "≈" reasonably accurate enough.

$$\frac{1}{\sqrt{\lambda_0}} \leftrightarrow 3.13 \cdot b - \frac{1.56 \cdot b^2}{a} \tag{3}$$

$$\frac{1}{\sqrt{\lambda_1}} \approx \underbrace{-2 \cdot \log_{10}\left(\frac{2.51}{Re} \cdot \left(3.13 \cdot b - \frac{1.56 \cdot b^2}{a}\right) + \frac{\varepsilon}{3.71 \cdot D}\right)}_{accelerated\ Eq.(3)} \tag{3a}$$

$$\frac{1}{\sqrt{\lambda_0}} \leftrightarrow b + 0.904 \cdot a + 1.08 \cdot \sin(0.937 \cdot a - b) - 1.85 \tag{4}$$

$$\frac{1}{\sqrt{\lambda_1}} \approx \underbrace{-2 \cdot \log_{10}\left(\frac{2.51}{Re} \cdot (b + 0.904 \cdot a + 1.08 \cdot \sin(0.937 \cdot a - b) - 1.85) + \frac{\varepsilon}{3.71 \cdot D}\right)}_{accelerated\ Eq.(4)} \tag{4a}$$

$$\alpha = \frac{1}{\sqrt{\lambda_0}} \approx a + 0.61 \cdot b + 0.28 \cdot a \cdot b + 0.51 \cdot sin(0.935 \cdot a - b) - 0.894 - 0.103 \cdot a^2 - 0.158 \cdot b^2 \tag{5}$$

$$\frac{1}{\sqrt{\lambda_1}} \approx \underbrace{-2 \cdot \log_{10}\left(\frac{2.51 \cdot \alpha}{Re} + \frac{\varepsilon}{3.71 \cdot D}\right)}_{accelerated\ Eq.(5)} \tag{5a}$$

$$\beta = \frac{1}{\sqrt{\lambda_0}} \approx 1.15 \cdot a + 0.569 \cdot b + 0.292 \cdot a \cdot b + 0.478 \cdot \sin(0.939 \cdot a - b) +$$
$$+ 0.122 \cdot \sin^2(0.939 \cdot a - b) - 1.284 - 0.12 \cdot a^2 - 0.162 \cdot b^2 \tag{6}$$

$$\frac{1}{\sqrt{\lambda_1}} \approx \underbrace{-2 \cdot \log_{10}\left(\frac{2.51 \cdot \beta}{Re} + \frac{\varepsilon}{3.71 \cdot D}\right)}_{accelerated\ Eq.(6)} \tag{6a}$$

Distribution of the relative error over the domain of applicability of the Colebrook equation introduced by these four approximations is in Figures 4-7.

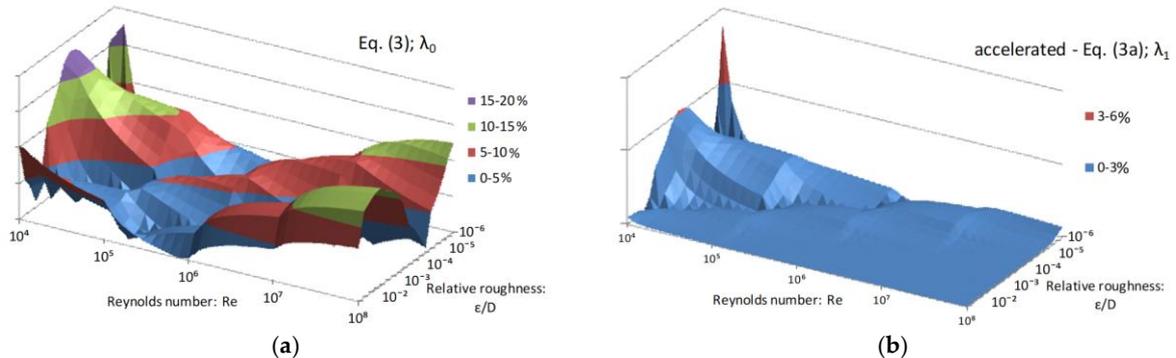

**Figure 4.** Distribution of the relative error of λ over the domain of applicability of the Colebrook equation introduced by Eq. (3) – left, and accelerated Eq. (3a) - right; relative error up to 20% and up to 5.35% respectively

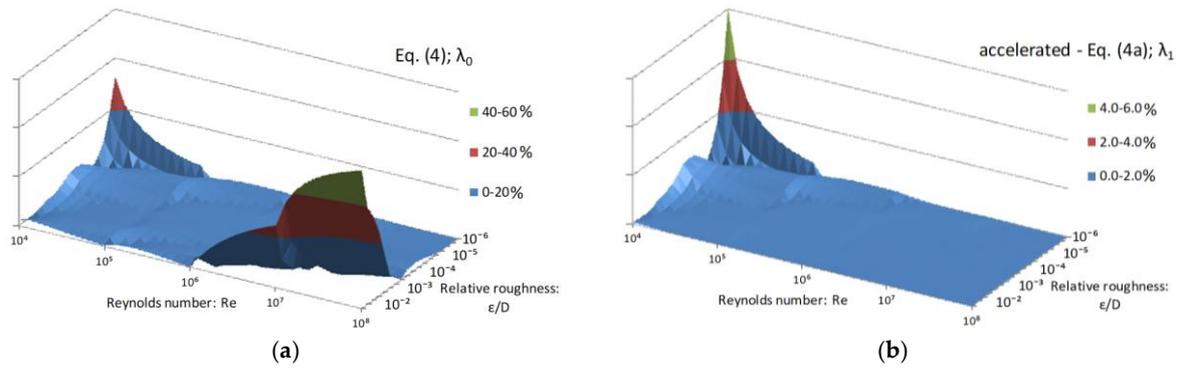

(**a**)            (**b**)

**Figure 5.** Distribution of the relative error of $\lambda$ over the domain of applicability of the Colebrook equation introduced by Eq. (4) – left, and accelerated Eq. (4a) - right; relative error up to 60% and up to 6.29% respectively

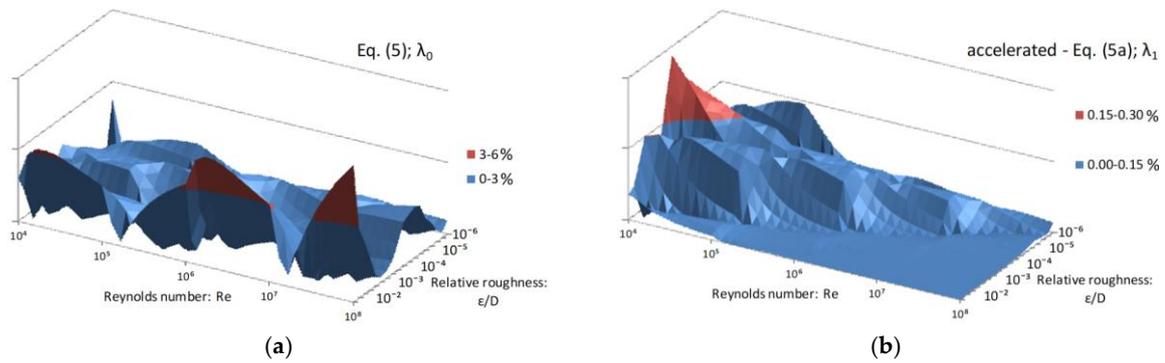

(**a**)            (**b**)

**Figure 6.** Distribution of the relative error of $\lambda$ over the domain of applicability of the Colebrook equation introduced by Eq. (5) – left, and accelerated Eq. (5a) - right; relative error up to 6% and up to 0.28% respectively

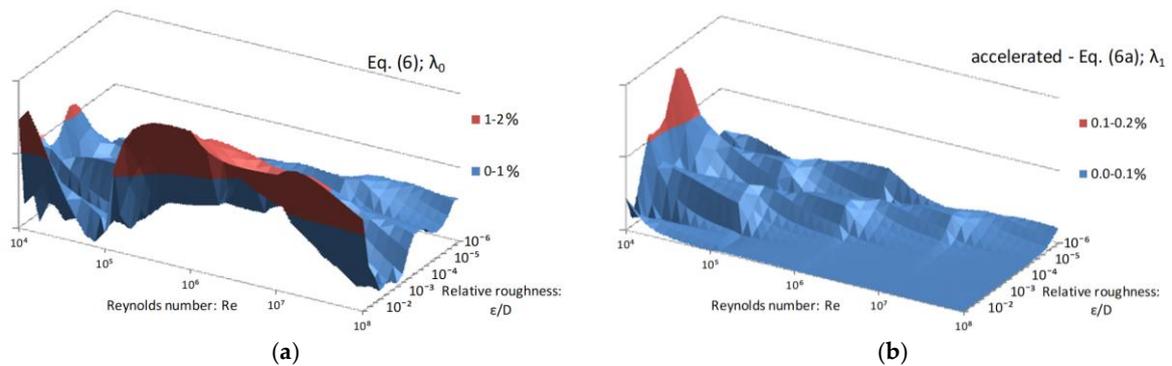

(**a**)            (**b**)

**Figure 7.** Distribution of the relative error of $\lambda$ over the domain of applicability of the Colebrook equation introduced by Eq. (6) – left, and accelerated Eq. (6a) - right; relative error up to 2% and up to 0.17% respectively

*2.3. Discussion – Comparative analysis*

In general, the relative error of approximations of the Colebrook equation is not uniformly dispersed over the domain of applicability of the Colebrook equation [7, 29], where the more complex is not always more accurate which can be noticed by comparing Eq. (3) and Eq. (4) and related Figures 4 and 5 respectively. According to Eureqa, the software package which generated the approximations, Eq. (5) is three times more complex compared with Eq. (3), while Eq. (4) 1.5 times more complex compared with Eq. (3). Using genetic algorithms [25, 29] or Excel fitting tool [27], there is a possibility to increase accuracy of the presented approximations by changing their parameters near the current value. If the structure of the approximations remains unchanged, the complexity remains also unchanged while accuracy can increase [4, 12, 13], or in other words the structure of the equations remains unchanged while the values of coefficient changes in order to fit better values obtained through the Colebrook equation. In that way error from Figures 4-7 can decrease but also distribution of the error over the domain of applicability changes [7, 29].

To summarize our findings, we made Table 1, in which the maximum relative errors of approximations compared to the accurate $\lambda$ are expressed as a function of complexity.

Table 1. The maximal relative errors of approximations as a function of complexity

| Accuracy | Complexity: Number of log functions | | |
|---|---|---|---|
| | 1 | 2 | 3 |
| High | | **Eq. (2a) – $\lambda_2 \rightarrow$0.13%** | Eq. (6a) - $\lambda_1 \rightarrow$0.17% |
| | | | Eq. (5a) - $\lambda_1 \rightarrow$0.28% |
| Moderate | Eq. (2a) - $\lambda_1 \rightarrow$0.98% | Eq. (6) – $\lambda_0 \rightarrow$2% | |
| Low | | Eq. (5) – $\lambda_0 \rightarrow$6% | Eq. (4a) - $\lambda_1 \rightarrow$6.29% |
| | | | Eq. (3a) - $\lambda_1 \rightarrow$5.35% |

It is clear that polynomial approximation accelerated through the fixed-point iterative procedure; Eq. (2a) shows better performances compared with those with normalized input parameters; Eqs. (3a-6a). For example, Eq. (2a) with two logarithmic functions (used for acceleration) gives relative error of no more than $\lambda_2 \rightarrow$0.13% compared with the approximately same error of Eq. (6a) with three logarithmic forms (two for normalization and one for acceleration); Eq. (6a) - $\lambda_1 \rightarrow$0.17%. Also, expression for $\lambda_0$ in case of Eq. (2) is polynomial while for Eqs. (4-6) contains also sinus trigonometric function [46]. After this careful analysis it can be concluded that it is better to use computationally expensive logarithmic functions for acceleration through Eq. (2a) and not for normalization.

All approximations can be classified as highly accurate, moderate and with low level of accuracy:
- Highly accurate: Compared with the similar approximations to the Colebrook equation, accelerated Eq. (5a); with the relative error up to 0.28% and accelerated Eq. (2a) and (6a) with the relative error up to 0.13% and 0.17%, respectively, are accurate as approximations by Bar [47] (0.2%), Chen [38] (0.36%-0.18%), Zigrang and Sylvester [41] (0.14%-0.08%, simpler: 1%-0.775%), Fang et al. [48] (0.61%-0.56%), Serghides [38] (0.14%-0.0026%, simpler 0.35%-0.27%), Buzzelli [49] (0.14%-0.08%), Sonad and Goudar [50] (0.8%-improved by Vatankhah and Kouchakzadeh [51]: 0.15%) and Romeo et al. [52] (0.14%-0.008%); where the higher reported accuracy is achieved through genetic optimization [25, 29]. These approximations are among the most accurate available until date [3-8], but at the same time in many cases much more complex compared to the approximations presented in our paper [4, 11-13]. For example; approximations by Barr [47] and by Chen [38] contain two logarithmic expressions and two non-integer powers; by Romeo et al. [52], three logarithmic expressions and two non-integer powers, etc. which means that they introduce a higher computational burden to achieve the same accuracy. In this case our Eq. (2a) after two steps of acceleration, which contains only two logarithmic forms.
- Moderately accurate: Our Eq. (6) with the relative error up to 2% does contain only two logarithmic expressions used for normalization and none non-integer power, and its accuracy can be compared with approximations by Swamee and Jain [53] (2.18%-1.75%), Manadili [54] (2%-1.5%), Brkić [42, 55-57] (2%-1.3%), Haland [58] (1.4%-1.1%), etc, all with the same or higher complexity as Eq. (6). Eq. (2a) after first step of acceleration with only one logarithmic function and with the relative error of up to 2.6% is even more efficient.
- Low accuracy: Our accelerated Eq. (3a) is very simple with the relative error up to 5.35% but with only one peak of high error (otherwise up to 3% as can be seen from Figure 4); it is more accurate compared with approximations by Round [59] (10.9%-5.5%), Eck [60] (8.2%-5.7%) and Avci and Karagoz [61] (4.8%-3.1%), Wood [62] (23.7%-16.6%), Moody [63] (21.5%-18.1%), etc.

## 3. Possible simplifications

As already noted, the main goal is to produce not only accurate, but also computationally low cost [11-13, 29] explicit approximations of the Colebrook equation. Trigonometric functions are used in Eqs. (4-6) and in their accelerated versions Eqs (4a-6a), which is not in common use related to the approximations of the Colebrook equation. These trigonometric functions can also have a higher computational cost. On the other hand, use of the Padé approximation $sin(x) \approx x \cdot (60-7 \cdot x^2)/(60+3 \cdot x^2)$ can potentially overwhelm the problem [64]. The Padé approximation $x \cdot (60-7 \cdot x^2)/(60+3 \cdot x^2)$ within the domain of interest; -0.08821<x<1.18456; compared with $sin(x)$ can introduce the relative error up to 0.068%. Within the same domain, Eureqa gives a number of approximations for $sin(x)$; but we have chosen to present here one accurate but relatively simple; $sin(x) \approx x-x^2/5350.6747-x^3/6.0171+x^5/127.4678$ with the relative error up to 0.003%. This error of approximations for $sin(x)$ also have impact on the final error of our approximations; Eqs. (4-6) and their accelerated pairs; Eqs. (4a-6a).

Also the Colebrook equation can be transformed in the mathematically equivalent form; Eq. (7) that is more suitable for further Padé simplifications. The main idea is to use already computed parameter $b=-log10(\varepsilon/D)$ and to use Padé polynomial in from $ln(1-\theta)$, where $\theta$ is given by Eq. (8). In Eq. (7), both $2 \cdot log_{10}(3.71) \approx 1.1387478$ and $2/ln(10) \approx 0.8686$ are constant, while $b=log_{10}(\varepsilon/D)$ is recycled as already evaluated during the normalization of input parameters.

$$\frac{1}{\sqrt{\lambda}} = 2 \cdot \log_{10}(3.71) - 2 \cdot \log_{10}\left(\frac{\varepsilon}{D}\right) - \frac{2}{\ln(10)} \cdot \ln\left(1 - \frac{-2.51 \cdot 3.71}{\frac{\varepsilon}{D} \cdot Re} \cdot \frac{1}{\sqrt{\lambda}}\right) \tag{7}$$

Note that Eq. (7) does not work for $\varepsilon/D=0$, but Eq. (1) does work; for practical application in case of smooth regime, both relations works; for example even for very smooth surfaces as for $\varepsilon/D=10^{-9}$. In practice [20], pipes are never that smooth to reach the value $\varepsilon/D=0$.

Our proposed acceleration through Eq. (2a) and Eqs. (3a-6a), can go further through Eq. (7a), where $b=-log_{10}(\varepsilon/D)$ and $\lambda_0$ from Eqs. (2-6).

$$\frac{1}{\sqrt{\lambda_1}} = 1.1387478 + 2 \cdot b - 0.8686 \cdot \ln\left(1 - \frac{-2.51 \cdot 3.71}{\frac{\varepsilon}{D} \cdot Re} \cdot \frac{1}{\sqrt{\lambda_0}}\right) \tag{7a}$$

The idea to eliminate the remained logarithmic form from the accelerated equations using Padé approximation [64] for $ln(1-\theta)$ where $\theta$ is defined by Eq. (8) in this case is not sustainable because of a wide domain of $\theta$ that contains a value between -30394.85651 and $-2.92065 \cdot 10^{-6}$ within the practical domain of the Colebrook equation; $4000<Re<10^8$ and $0<\varepsilon/D<0.05$.

$$\theta = \frac{-2.51 \cdot 3.71}{\frac{\varepsilon}{D} \cdot Re} \cdot \frac{1}{\sqrt{\lambda_0}} \tag{8}$$

Regarding the normalization of the input parameter $a=log_{10}(Re)$, we hoped also to use the fact that the practical domain of the Reynolds number Re, is from 4000 to $10^8$ which has as a consequence $log_{10}(Re) \approx log10(1+Re)$; where for $ln(1+y)$ numerous approximations are available, but mostly for the argument z around 0. This is because some computer algebra systems and programming languages provide a special natural logarithm plus 1 function alternatively named to give more accurate results for values of x close to zero compared to using $ln(1+y)$ directly. In our case this is not of interest, knowing that for $y=Re$; $y \gg 1$. Of course, to use only numbers between 1 and 10, we can use the rule $ln(Re)=ln(z \cdot 10^n)=ln(z)+n \cdot ln(10)$ where $n=len(int(z))$ and $z=Re/10^n$; len is a function which calculates number of digits in a number while int is function which gives a number down to the nearest integer; $ln(10)=2.30258509$. Similar can be done for the normalized parameter $b=-log_{10}(\varepsilon/D)$ where $\varepsilon/D$ is between 0 and 0.05.

A fast but still reliable Padé approximation useful for the Colebrook equation has been recently introduced in [10]. The logarithm term $log_{10}(z) = \frac{ln(z)}{ln(10)}$, where argument $z \sim 1$ is approximated by the rational function:

$$ln(z) \approx \frac{z \cdot (z \cdot (11 \cdot z + 27) - 27) - 11}{z \cdot (z \cdot (3 \cdot z + 27) + 27) + 3} \tag{9}$$

In this case, only one logarithm is computed and stored in the computer memory: $log_{10}(y_1)$, whereas $log_{10}(y_{i+1})$ is approximated by $log_{10}(y_1)$ and by Eq. (9) as

$$log_{10}(y_{1+1}) = log_{10}(y_1) - log_{10}(z), \qquad (10)$$

where $z = \frac{y_1}{y_{i+1}}$ is defined in Eq. (2b). When the Padé approximation (9) is applied to Eq. (2a) and the starting point is estimated by Eq. (2), the maximum relative error of Eq. (10) is negligible: 5·10^{-10}%. Thus, Eq. (10) is very accurate, when the iterative process defined by Eq. (2a) is initiated by a good starting point (A comparison of iterative methods for solving the Colebrook equation is given in [65]). Consequently, the argument $z$ of Eq. (9) is very close to one in all cases within the domain of interest of the Colebrook equation [10].

**4. Conclusion**

Evaluation of hydraulic resistance, i.e. computation of flow friction factor, $\lambda$ is one of the main tasks encountered in engineering practice wherever flow of fluid through closed conduits occur [43, 44]. Up to now, the empirical Colebrook Equation (1) which is implicitly given by a flow friction factor, $\lambda$ is still accepted as an informal standard after 80 years. Obvious disadvantages of an implicit relationship inspired by numerous efforts to derive as accurate possible explicit equivalent; $\lambda \approx f(Re, \varepsilon/D)$ to the original Colebrook equation; $\lambda=f(Re, \varepsilon/D, \lambda)$ [3-8, 65]. Nowadays, the requirement is not only to develop accurate, but also computationally efficient [11-13] approximations to the Colebrook equation inspired us to use the ability of artificial intelligence to recognize patterns not knowing their true physical laws [15, 29]. Using genetic programming we developed a few accurate approximations to the Colebrook equation avoiding extensive use of computationally expensive logarithmic forms [11-13] on which Colebrook's relation is based. The most accurate approximations here presented are with the relative error of up to 0.13% with only two and three logarithmic forms used which makes it very balanced in ratio between accuracy and computational efficiency []. The polynomial expression; Eq. (2) accelerated through the two steps of fixed-point iterative procedure; Eq. (2a) introduced the relative error up to 0.13% using only two logarithmic functions. Practically the same error is introduced, when only one logarithmic function is computed, whereas the second logarithm is approximated by the Padé approximation. On the other hand, Eq. (6a) reaches approximately the same accuracy using three logarithmic functions (two for normalization of input parameters and one for fixed-point acceleration) and using also one sinus trigonometric function. Therefore, a polynomial function can be recommended because it is cheaper for computation and because acceleration through the fixed-point iterative procedure is more efficient compared with the normalization of input parameters [15].

Our genetic approximations are valid only for a turbulent regime, i.e. for a simulation of the Colebrook equation. A transition from a laminar to a turbulent regime is rapid and it is not covered by the Colebrook formula. Our previous work with artificial neural networks [15] shows that this transition cannot be simulated easily using artificial intelligence techniques. We found [15] that the Colebrook equation can be simulated extremely accurately by a neural network with 50 neurons in the hidden layer. However, with the transition from laminar to turbulent regime included, even such complex network with 50 neurons introduces a very high error in this critical zone even after long lasted and complex strategies of training. Similar findings are also valid for the genetic approach.